\documentclass[prd,twocolumn,amsmath,amssymb,floatfix,superscriptaddress]{revtex4}
 \usepackage{graphicx}
 \usepackage{amsmath,amssymb}
 \usepackage{bm}
 \usepackage{verbatim,color}

\DeclareFontFamily{OT1}{pzc}{}
\DeclareFontShape{OT1}{pzc}{m}{it}%
             {<-> s * [1.00] pzcmi7t}{}
\DeclareMathAlphabet{\mathscr}{OT1}{pzc}%
                                 {m}{it}

\newcommand{\be}{\begin{equation}}
\newcommand{\ee}{\end{equation}}
\newcommand{\bea}{\begin{eqnarray}}
\newcommand{\eea}{\end{eqnarray}}
\newcommand{\refeq}[1]{Eq.~(\ref{eqn:#1})}

\newcommand{\reffig}[1]{Fig.~\ref{fig:#1}}          
\newcommand{\refFig}[1]{Figure~\ref{fig:#1}}
\newcommand{\reftab}[1]{Tab.~\ref{t:#1}}
\newcommand{\vs}{\nonumber\\}       
\newcommand{\refsec}[1]{Sec.~\ref{sec:#1}}          
\def\ba#1\ea{\begin{align}#1\end{align}}

   \newcommand{\planck}{Planck}
   \newcommand{\fnl}{f_{\rm NL}}

\begin{document}
 
   \title{Constraints on Non-Gaussianity from Sunyaev--Zeldovich Cluster Surveys}
 
 \author{Daisy S. Y. Mak}
 \email{suetyinm@usc.edu}
\affiliation{Physics and Astronomy Department, University of Southern California, Los Angeles, California 90089-0484, USA}

\author{Elena Pierpaoli}
 \email{pierpaol@usc.edu}
\affiliation{Physics and Astronomy Department, University of Southern California, Los Angeles, California 90089-0484, USA}

\begin{abstract}
We perform a Fisher matrix analysis to forecast the capability of ongoing and future Sunyaev-Zeldovich cluster surveys in constraining the deviations from Gaussian distribution of primordial density perturbations. We use the constraining power of the cluster number counts and clustering properties to forecast limits on the $\fnl$ parameter. The primordial non-Gaussianity effects on the mass function and halo bias are considered. We adopt self-calibration for the mass-observable scaling relation, and evaluate constraints for the SPT, Planck, CCAT--like, SPTPol and ACTPol surveys.   We show that the scale-dependence of halo bias induced by the local NG provides strong constraints on $\fnl$, while the results from number count are two orders of magnitude worse. When combining information from number counts and power spectrum, the \planck\ cluster catalog provides the tightest constraint with $\sigma_{\fnl}=7$ ($68\%$ C.L.) even for relatively conservative assumptions on the expected cluster yields and systematics. This value is a factor of 2 smaller than the $1\sigma$ error as measured by WMAP CMB measurements, and comparable to what expected from Planck. 
We find that the results are mildly sensitive to the mass threshold of the surveys, but strongly depend  on the survey coverage: a full-sky survey like Planck is more favorable because it can probe longer wavelengths modes which are most sensitive to NG effects. In addition, the constraints are largely insensitive to priors on nuisance parameters as they are mainly driven by the power spectrum probe which has a mild dependence on the mass-observable relations. 

\end{abstract}

\maketitle

\section{Introduction}
Cosmological inflation has emerged as the most popular scenario of the early universe and it predicts a near scale invariant power spectrum and close--to Gaussian distribution for the primordial curvature inhomogeneities that seeds large scale structure (LSS). Various inflationary models produce different levels of departures from Gaussianity. For example, the  slow--roll and single field inflation model produces tiny amount of departures from Gaussianity, while other models predict sizable amount of primordial Non-Gaussianity (NG) that could be observed with current experiments such as CMB and galaxy clusters~\cite[e.g.][]{Bartolo2004,Chen2010}. Therefore, any detection of primordial NG would open a new and extremely informative window on the physics of inflation and the very early Universe. 

Current measurements of the CMB~\cite[e.g.][]{Komatsu2011} and LSS~\cite[e.g.][]{Slosar2008} found that the distribution of primordial fluctuations is consistent with Gaussianity, however, that bound is still several orders of magnitude away from testing primordial NG at the level predicted by slow--roll inflation. Galaxy clusters are in principle sensitive and powerful tool for this purpose because they trace the rare, high mass tail of density perturbations. As a result, the changes in the shape and evolution of the mass function of dark matter halos are most sensitive to departures from Gaussianity. The effect of NG on the mass function has been investigated in e.g.~\cite{Matarrese2000,Mathis2004,Grossi2007,Kang2007,Maggiore2010,Sefusatti2007} and was validated by large-scale cosmological simulations with non-Gaussian initial conditions~\cite{Grossi2007,Grossi2009,Dalal2008,Desjacques2009}. More recently NG effects on the large scale clustering of collapsed haloes were studied by~\cite{Dalal2008,Matarrese2008,Valageas2010,Lam2009}. They found that the linear biasing parameter acquires a scale dependence, which modifies the power spectrum of the the distribution of cosmic structures most prominently at large scales. This unique signature serves as a powerful way to constrain and forecast the nature of NG assumption. For instance, predictions for the cluster abundance and clustering of galaxy clusters expected from various surveys with primordial NG conditions were presented in~\cite{Fedeli2009,Roncarelli2010}.

In this work, we aim to forecast the capability of ongoing and future SZ cluster surveys in constraining primordial NG, using the Fisher Matrix approach. Several initial studies have explored such possibility with future galaxy cluster surveys in different wavebands, e.g. X-ray~\cite{Sartoris2012,Pillepich2012} and optical~\cite{Fedeli2011,Cunha2010,Oguri2009}, and they all showed that the constraints from galaxy clusters are quite strong. With the use of SZ cluster survey, we would complete the spectrum of cosmological applications with galaxy clusters. Forthcoming SZ experiments will provide large samples of mass selected clusters, and multi--frequency followup observations can add information on the cluster mass. In addition, these surveys will detect clusters at high redshift to test non-Gaussianity in the regimes where its effects are pronounced, that is, the high mass tail of the mass function and the large scale power spectrum of the cluster distribution. This makes the SZ clusters more favorable since X--ray and optical clusters are not as efficient in detecting high redshift clusters.

This paper is organized as follows. We begin by presenting the surveys and expected cluster samples in \refsec{data}. In \refsec{theory} we present the parametrization of primordial non-Gaussianity effects on the halo abundance and clustering. \refsec{fisher} details the Fisher formalism employed here, as well as the fiducial cosmology adopted. The forecasted constraints are presented in \refsec{result}. We discuss our results in \refsec{discussion} and conclude in \refsec{conclusion}.

\section{Cluster surveys}
\label{sec:data}

\begin{figure}
  \begin{center}
\includegraphics[width=0.50\textwidth]{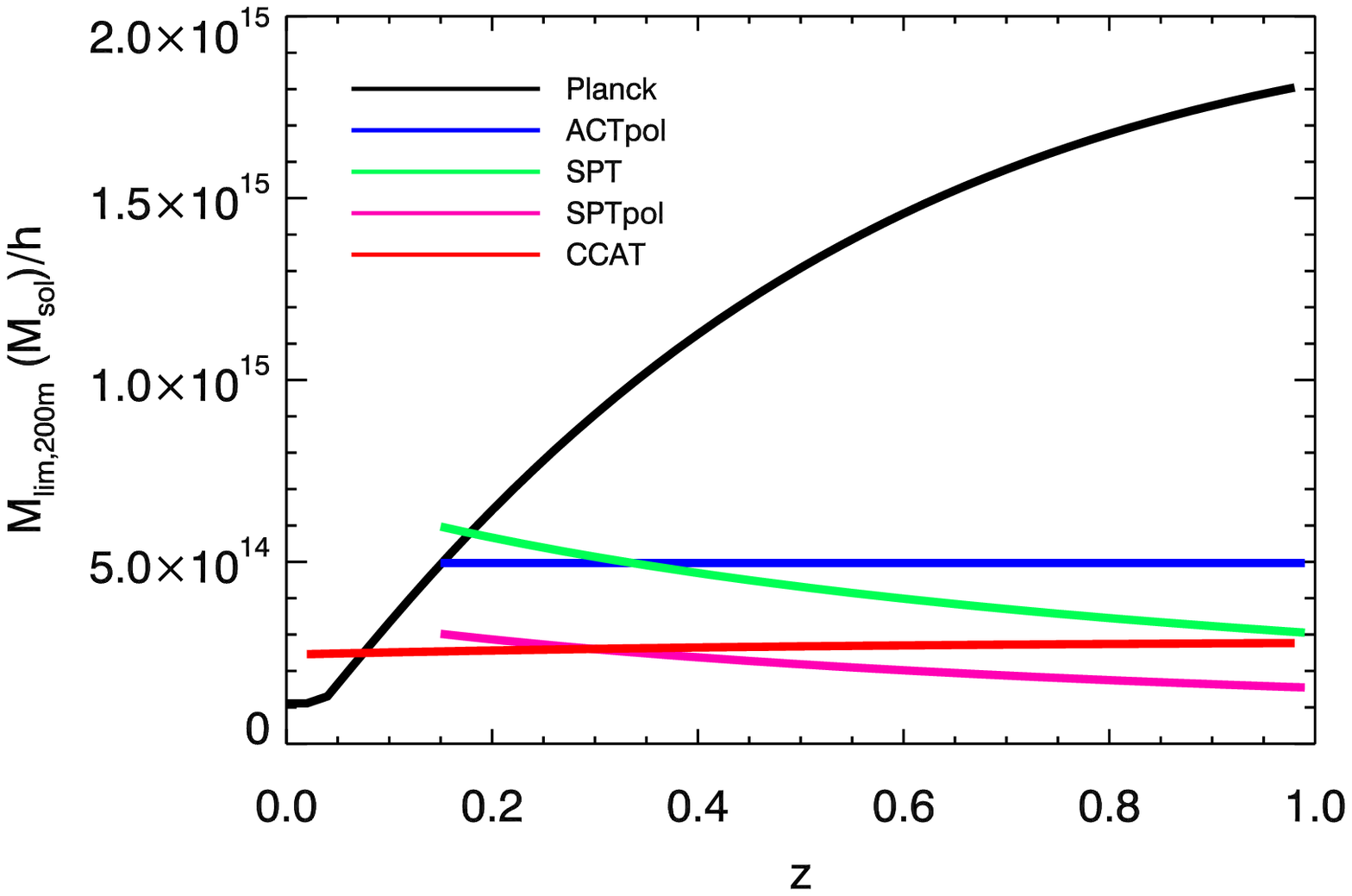}  
\includegraphics[width=0.50\textwidth]{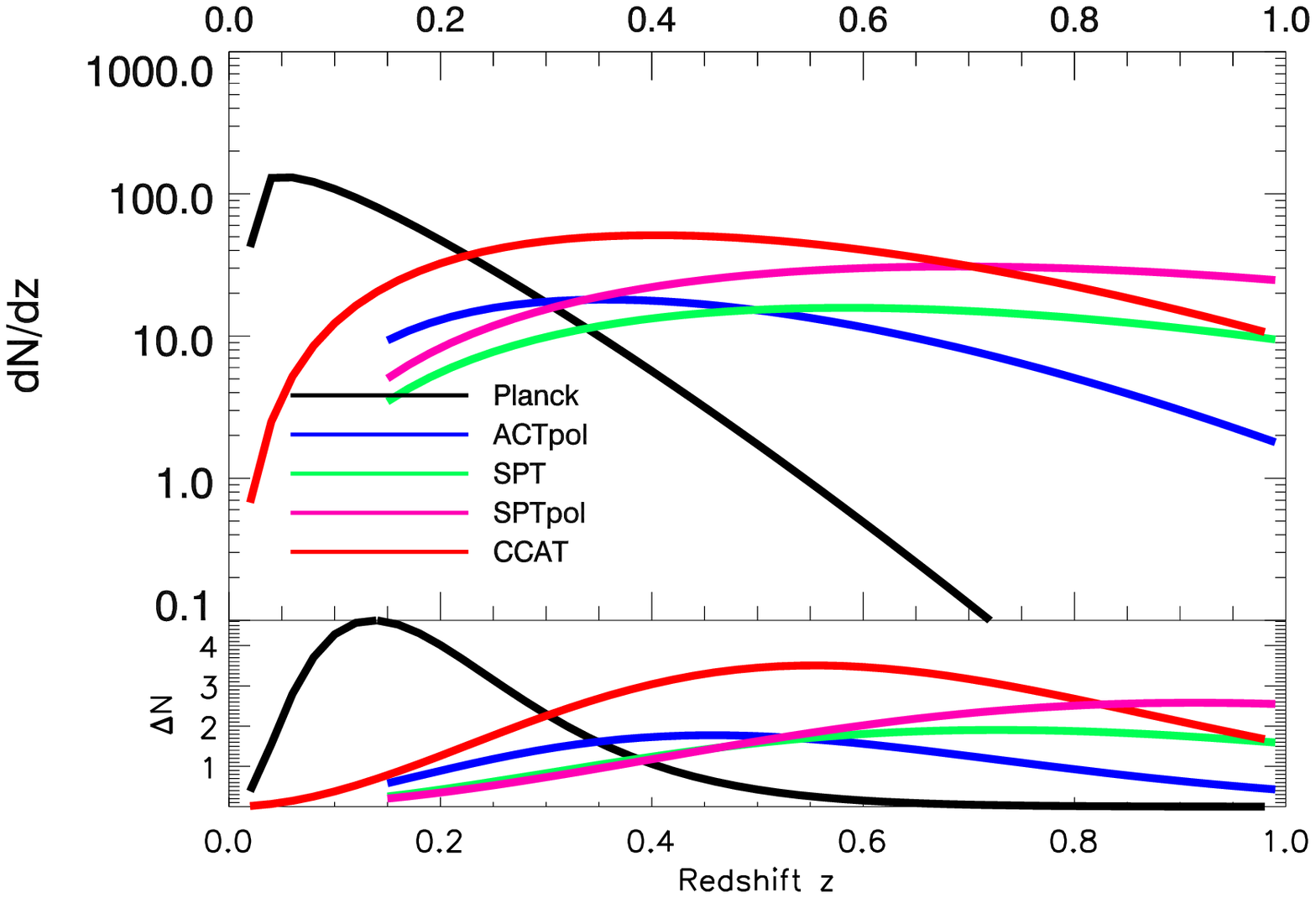} 
         \caption{{\it Upper:} Mass limit of cluster surveys. {\it Lower:} The redshift distribution of clusters in the Planck (black), ACTpol (blue), CCAT--like (red), SPT (green), and SPTpol (magenta) survey in the fiducial $\Lambda$CDM cosmology (solid). The bottom panel shows the fraction deviation from the gaussian number count when $\fnl=100$.}
     \label{fig:Nz}
  \end{center}
\end{figure}

We will investigate the predictions for the five surveys described below.  While we try to obtain as realistic survey specifications as possible, in particular for the mass limit as function of redshift 
$M_{\rm lim}(z)$, the intrinsic scatter in the mass observation relations and the lack of previous large samples of SZ clusters necessarily make these quantities somewhat uncertain.  In particular,
the relation between cluster mass and SZ signal is still very uncertain
(e.g.~\cite{Ameglio2009,Rasia2005,Nagai2007,PifVal2008}).  
The final mass limits as a function of redshift are shown in the upper panel of \reffig{Nz}, and the resulting expected number of clusters for each survey is shown in the lower panel of \reffig{Nz}.  
These limits are derived considering the characteristics of each survey, as specified in the following subsections.

\subsection{The \planck\ Catalog} 
\planck\ is imaging the whole sky with an unprecedented combination of sensitivity ($\Delta T/T\sim2\times10^{-6}$ per beam  at 100 - 217 GHz), angular resolution ($5'$ at 217 GHz), and frequency coverage ($30-857$ GHz). The SZ signal is expected to be detected from a few thousand individual galaxy clusters.  \planck\ will produce a cluster sample with median redshift $\sim0.3$ (see \reffig{Nz}, upper left panel). The SZ observable is the integrated Comptonization parameter $Y=\int y\ d\Omega_{\rm cluster}$  out to a given radius. For Planck,  a $5\sigma$  detection threshold ensuring high level of completeness (about 90\%)  corresponds to  $Y_{200,\rho_c}\ge2\times10^{-3}{\rm arcmin^2}$~\citep{Melin2006}, where $Y_{200,\rho_c}$ is 
the integrated comptonization parameter within $r_{200,\rho_c}$, the
radius enclosing a mean density of 200 times the critical density.  
The early release from the \planck\ Collaboration gives a sample of 189 high signal-to-noise SZ clusters with $\ge6\sigma$ detection. 
 It is therefore likely that our assumed detection threshold  will be 
 eventually reached in future data releases.
  For an SZ survey, its flux limit can be translated into a limiting mass 
by using simulation-calibrated scaling relations \cite{Sehgal2007}:

\begin{equation}
\frac{M_{\rm lim,200\rho_c}(z)}{10^{15}M_\odot } =\left[\left (\frac{D_A(z)}{{\rm Mpc}/h_{70}}\right)^2 E(z)^{-2/3} \frac{Y_{\rm 200, \rho_c}}{2.5\times10^{-4}} \right] ^{0.533}.
\label{eqn:szflux}
\end{equation}

In order to mitigate the effect of overestimation of unresolved clusters at low redshift, we further restrict $M_{\rm lim,200\rho_c}$ to be at least $10^{14}M_{\odot}$ at all $z$. 
With all these criteria, the \planck\ survey is expected to detect $\sim1000$ clusters.
The mass threshold we find with this approach is consistent with the one in \cite{Schafer2007}.
While we keep $Y_{\rm 200, \rho_c}=2\times10^{-3}{\rm arcmin^2}$ as our reference minimum value for presentation of the  main results, we will 
also discuss predictions for a lower mass threshold, corresponding to $Y_{\rm 200, \rho_c}=10^{-3}{\rm arcmin^2}$. With such threshold, the completeness of the  $S/N > 5$ sample is reduced to about 70\%
and the total number of clusters is 2700.

\subsection{SPT and SPTpol}
\label{sec:spt}
The SPT survey is currently observing the sky with a sensitivity of $18 \mu$K/arcmin$^2$ at 148 GHz, 218 GHz, and 277 GHz. This survey covers $\Omega\approx2500$ square degrees of the southern sky (between $20h\geq{\rm RA}\geq7h$, $-65^{\circ}\le\delta\le-30^{\circ}$) with a projected survey size and cluster mass limit well matched to the Stage III survey specification of the Dark Energy Task Force~\cite{Vanderlinde2010}. 
For the mass limits, we employ the calibrated selection function of the survey by~\cite{Vanderlinde2010}.  This is based on simulations and used to provide a realistic measure of the SPT detection significance and mass. Disregarding the scatter in the fitting parameters for this relation, we use here:
\be
\frac{M_{\rm lim,200\bar{\rho}}(z)}{5\times10^{14}M_{\odot}h^{-1}}=\left [ \left ( \frac{\sqrt{\xi^2-3}}{6.01} \right )\left ( \frac{1+z}{1.6} \right )^{-1.6} \right ]^{1/1.31}
\label{eqn:spt}
\ee

\noindent where $\xi$ is the detection significance. For the SPT survey, we take clusters detected at $\xi>5$ which ensure a 90\% purity level.  Currently, the SPT team is setting a low redshift cut at $z_{\rm cut}=0.3$ in 
their released cluster sample, due to difficulties in reliably distinguishing  low-redshift clusters from CMB fluctuations 
in single frequency observations. Nevertheless, with upcoming multi-frequency observations, a lower cut $z_{\rm cut}=0.15$ will likely be attained. We therefore apply this cut in our work. 
With this, the SPT survey is expected to detect $\sim500$ clusters.

In addition to this, we also consider the upcoming SPT polarization survey (hereafter SPTpol) which will have an increased sensitivity of $4.5\mu$K/arcmin$^2$ at 150 GHz for a 3 year survey and sky coverage of 625 square degrees.  We scaled the mass limits by a factor of $3.01/5.95$ in \refeq{spt} to match with the expected mass limits of SPTpol clusters (Benson 2011, private communication).  We again use $z_{\rm cut}=0.15$, resulting in a total expected number of $\sim1000$ clusters.  
While these are the limits we use for our main results, we also discuss outcomes that consider a lower mass limit, corresponding to $\xi = 4.5$ (80\% purity). With this mass limit, SPT would find 800 clusters and SPTPol about 1400 clusters.

\subsection{ACTpol}
The Atacama Cosmology Telescope (ACT) has been observing a portion of the southern sky since 2008 consisting of two strips of the sky, each 4 degrees wide in declination and 360 degrees around in right ascension, one strip is centered at $\delta=-5^{\circ}$, and the other is centered at $\delta=-55^{\circ}$~\cite{Sehgal2007}. With a sensitivity of $\approx35\mu$K/arcmin$^2$, only about 100 clusters are expected to be detected. Instead, we turn to the newly developing dual-frequency (150 GHz and 220 GHz) polarization sensitive receiver (hereafter ACTpol~\cite{Niemack2010} and reference therein) to be deployed on ACT in 2013. One of the three ACTpol observing seasons will have a wide survey covering $4000{\rm deg}^2$ to a target sensitivity of $20\mu$K/arcmin$^2$ in temperature at 150 GHz. With the wide field, they aim to find $\sim600$ clusters in the ACTpol survey.
The survey is $90\%$ complete above a limiting mass of $M_{\rm lim,200\bar{\rho}}=5\times10^{14}M_\odot h^{-1}$ (Sehgal 2011, private communication), and we therefore assume this as our redshift-independent mass limit for ACTpol.  As in SPT, the ACT team also put a low redshift cut in their parameter determination works and we likewise take $z_{\rm cut}=0.15$ for ACTpol, resulting in a total expected number of $\sim500$ clusters.  
We also present in the discussion section the results corresponding to a lower mass limit, 
$M_{\rm lim,200\bar{\rho}}=4\times10^{14}M_\odot h^{-1}$, which would result in a catalog of about 1000 clusters.

\subsection{CCAT--like}
The Cornell Caltech Atacama Telescope (CCAT--like) will be a 25 meter telescope for observations at submm wavelengths. It will combine high spatial resolution ($\approx5''$), a wide field of view ($20'$), and a broad wavelength range (100 GHz--405 GHz) to provide an unprecedented capability for deep, large area, multi-color SZ surveys of galaxy clusters to complement narrow field, high resolution studies with ALMA. The telescope aims for initial observations in 2015 and will be the first  to provide a large sample of clusters with high spatial resolution SZ profiles that aid in studying cluster astrophysics. 

We consider here a prototype of CCAT--like catalog from the CCAT--like LWcam survey which will cover 2000 square degrees of the sky for 1000 hours, resulting in a noise level of $12\mu K$/arcmin at 220 GHz. Following the LWcam Design Study Proposal (private communication, Vanderlinde 2012), we estimate its mass limits using the framework developed for the SPT (see~\refsec{spt}). Likewise, we consider clusters detected at $\xi>5$ (90\% purity level) in the redshift range $z=0-1$. Using~\refeq{spt}, the CCAT--like catalog is expected to detect 1500 clusters.

\section{Primordial Non-Gaussanity}
\label{sec:theory}
Early universe models predict deviation from Gaussian initial conditions. Primordial non-Gaussianity induced by inflationary models can be conveniently parametrized by a nonlinear coupling parameter $\fnl$ and the Bardeen's gauge invariant potential $\Phi$. This can be written as the sum of a linear Gaussian term and a non-linear second order term that encapsulates the deviation from Gaussianity~\cite{Salopek1990,Gangui1994,Verde2000,Komatsu2001}

\be
\Phi = \Phi_{\rm G} + \fnl (\Phi^2_{\rm G} - \left \langle\Phi_{\rm G}^2 \right \rangle)
\ee

The parameter $\fnl$ determines the amplitude of the non-Gaussianity. In this work, we adopt the large scale structure convention for defining the fundamental parameter $\fnl$. As noted by~\cite{Grossi2009}, the primordial value of $\Phi$ has to be linearly extrapolated at $z=0$ in the LSS convention, so that $\fnl=g(z)/g(0)f_{\rm NL}^{\rm CMB}\approx1.3f_{\rm NL}^{\rm CMB}$ for the $\Lambda$CDM model. We note that the factor 1.3 is approximate since the growth factor $g(z)$ is cosmological dependent. This means that any constraints gathered from CMB data should be increased by $30\%$ in order to comply with the convention adopted here.

We define here several terms which appear in subsequent text. The relation between the power spectrum matter density fluctuation extrapolated at $z=0$,  $P(\vec{k})$, and the power spectrum of the Newtonian potential, $P_\Phi(\vec{k})$ is

\be
P(\vec{k}) T^2(k) = \left [\frac{2T(k) k^2}{3H_0^2 \Omega_{m,0}} \right ]^2 P_\Phi(\vec{k}) = M^2_R(k) P_\Phi(\vec{k})
\ee

\noindent The primordial matter power spectrum is scale free, i.e. $P(\vec{k})=Ak^{n}$, and therefore the potential power spectrum can be written as $P_\Phi(\vec{k}) = \frac{9AH_0^4\Omega_{m,0}^2}{4} k^{n-4}  \equiv Bk^{n-4}$. 

In the case of non-Gaussianity, the random field of the potential $\Phi$ cannot be described by the power spectrum $P_{\Phi}=Bk^{n-4}$ alone. Higher order moments, in particular the bispectrum $B_{\Phi}(\vec{k_1},\vec{k_2},\vec{k_3})$, are required. The bispectrum is defined on the basis of the Fourier transform of the three point correlation function $\left \langle \Phi(\vec{k_1})\Phi(\vec{k_2})\Phi(\vec{k_3})\right \rangle$ as follows

\be
\left \langle \Phi(\vec{k_1})\Phi(\vec{k_2})\Phi(\vec{k_3})\right \rangle\equiv(2\pi)^3\delta_{\rm D}(\vec{k_1}+\vec{k_2}+\vec{k_3})B_\Phi(\vec{k_1},\vec{k_2},\vec{k_3})
\ee

The shape of the non-Gaussian bispectrum is related to the fundamental physics of the early universe and the evolution of the inflation field. A wide class of inflationary scenarios lead to non-Gaussianity of the local type in which the bispectrum of the Bardeen's potential is maximized for squeezed configurations. Example of this scenario is the curvaton model which involves an additional contribution to the curvature perturbations by a light field~\cite{Lyth2003}. The parameter $\fnl$ of the local type is a constant in space and time with $\fnl \ll1$, and is expected to be of the same order of the slow-roll parameters~\cite{Falk1993}. In such case, the bispectrum has a simple form~\cite{Catelan1995}:

\be
B_\Phi(\vec{k_1},\vec{k_2},\vec{k_3}) = 2\fnl B^2 (k_1^{n-4} k_2^{n-4} + k_1^{n-4} k_3^{n-4} + k_2^{n-4} k_3^{n-4}) 
\ee

\subsection{Mass Function}
Non-gaussianity lead to a modified mass function with respect to the gaussian case and are usually expressed as perturbations of the gaussian mass function. There are several prescriptions of the corrections in mass functions of collapsed objects (e.g.~\cite{Matarrese2000},~\cite{Loverde2008}). In this work, we adopt the approach of~\cite{Loverde2008} (hereafter LMVJ), in which the probability distribution for the smoothed dark matter density field is approximated using the Edgeworth expansion truncated at the the first few orders. The LMVJ approach was shown to give reasonable agreement with full numerical simulations of structure formation~\cite{Grossi2009}, provided that the linear overdensity threshold for collapse is corrected for ellipsoidal density perturbations, i.e. $\Delta_c\to\Delta_c\sqrt{q}$ where $q\approx0.75$.

In this prescription, the non-Gaussian mass function $n_{\rm NG}(M,z)$ can be written as a function of a Gaussian one, $n_{\rm G}$, multiplied by a non Gaussian correction factor $\mathit{R}(M,z)$, 
\be
n_{\rm NG}(M,z)=\mathit{R}(M,z) n_{\rm G}(M,z)
\label{eqn:nn}
\ee

\noindent where $\mathit{R}(M,z)\equiv n_{\rm G, PS}/n_{\rm NG, PS}$, and $n_{\rm G,PS}$ and $n_{\rm NG,PS}$ are the Gaussian and non-Gaussian mass function respectively computed according to the Press and Schechter formula~\cite{Press1974}. In this work, we adopt the formula from~\cite{Tinker2008} for the Gaussian mass function $n_{\rm G}(M,z)$. Using LMVJ approach, the correction factor $\mathit{R}$ is

\be
\mathit{R}(M,z) = 1 + \frac{1}{6}\frac{\sigma^2_{\rm M}}{\delta_c} \left [ S_3 (\frac{\delta_{\rm c}^4}{\sigma^4_{\rm M}} - 2\frac{\delta_{\rm c}^2}{\sigma^2_{\rm M}}-1) + \frac{dS_3}{d\ln \sigma_{\rm M}} (\frac{\delta_{\rm c}^2}{\sigma^2_{\rm M}}-1) \right ]
\label{eq:R}
\ee

\noindent where $\delta_c\equiv\Delta_c/D(z)$ is the critical density for collapse, $D(z)$ is the linear growth factor, $\sigma_{\rm M}$ is the rms of primordial density fluctuations on the scale corresponding to mass $M$, $S_3\equiv f_{\rm NL} \mu_3(M)/\sigma^4_{\rm M}$ is the normalized skewness, $\mu_3$ is the third order moment. For the local non-Gaussianity, $\mu_3$ can be computed as

\bea
& \mu_3(M)=  \frac{f_{\rm NL}}{(2\pi^2)^3} \int_0^{\infty}{\frac{dk_1}{k_1} M_R(k_1) P(k_1) \int_0^\infty{\frac{dk_2}{k_2} M_R(k_2) P(k_2) }} \nonumber \vs
&\int_{-1}^1{d\mu M_R(k_{12}) \left [ 1 + 2\frac{P(k_{12})}{P(k_2)} \right ] } 
\label{eq:mu3}
\eea

\noindent where $k_{12}^2=k_1^2 + k_2^2 + 2\mu k_1 k_2$. This integral is computational intensive. In order to reduce the workload of this calculation, we instead use the fitting formula of $S_3$ by~\cite{Chongchitnan2010} which is shown to have sub-percent accuracy:

\be
S_3=\frac{3.15\times10^{-4}f_{\rm NL}}{\sigma_R^{0.838}}
\label{eq:fittings3}
\ee

\noindent We verified the accuracy of this fitting formula by directly comparing this with the numerical table publicly available online~\footnote{http://icc.ub.edu/~liciaverde/nongaussian.html}, and we found that they agree with each other to percent level.

\subsection{Halo Bias}
The halo bias acquires an extra scale dependence due to primordial non-Gaussianity of the local type~\cite{Matarrese2008}, i.e. 

\be
b_{\rm NG}(M,z,k) = b_{\rm G}(M,z) + \Delta b(M,z,k)
\ee

\noindent where

\be
\Delta b(M,z,k) = [b_{\rm G}(M,z) -1 ] \delta_c \Gamma_R(k)
\ee

The term $\Gamma_R(k)$ encapsulates the dependence on the scale and mass. For the local non-Gaussianity it can be written as

\bea
& \Gamma_R(k)=\frac{2f_{\rm NL}}{8\pi^2 M_R(k) \sigma^2_R} \int_0^\infty{d k_1 k_1^2 M_R(k_1) P(k_1)} \vs
& \int_{-1}^1 {d\mu M_R(k_{12}) \left[ \frac{P(k_{12})}{P(k)} +2 \right]}
\eea

\noindent where $k_{12}^2=k_1^2 + k^2 + 2\mu k_1 k$. The function $\Gamma_R(k)$ is flat at small scales but scales as $k^{-2}$ at large scales ($k\ge0.01$ Mpc/h), so that a substantial deviation in the halo bias is expected at those scales. We adopt the formula from~\cite{Sheth2001} for the Gaussian bias $b_{\rm G}$.

\section{Fisher Matrix forecast}
\label{sec:fisher}
The Fisher information Matrix (FM hereafter) is defined as

\be
F_{\alpha\beta}\equiv -\left \langle \frac{\partial^2 \ln\mathfrak{L}}{\partial p_{\alpha} p_{\beta}} \right \rangle
\label{eqn:fisher}
\ee

\noindent where $\mathfrak{L}$ is the likelihood of a data set,
e.g. a cluster sample, written as a function of the parameters $p_\alpha$ describing
the model.  The parameters $p_\alpha$ comprise the cosmological model
parameters as well as ``nuisance'' parameters related to the data set
(e.g., mass calibration).  

\subsection{Cosmological parameters}

Throughout this paper, we assume a spatially flat ($\Omega_k=0$) cosmology.  
Our model comprises a total of seven cosmological parameters and the non-gaussianity $\fnl$ parameter which are left free to vary.  The seven parameters and their
fiducial values (in parenthesis, taken from the best-fit flat $\Lambda$CDM model from  
WMAP 7yr data, BAO and $H_0$ measurements \cite{Komatsu2011}) are: 
baryon density parameter $\Omega_bh^2$(0.0245); matter density parameter
$\omega_m \equiv \Omega_m h^2$ (0.143);
dark energy density $\Omega_\Lambda=1-\Omega_m$ (0.73);  power spectrum 
normalization $\sigma_8$ (0.809); index of power spectrum $n_s$ (0.963); 
effective dark energy equation of state through 
$w(z)=w_0+(1-a)w_a$, with fiducial values $w_0=-1$ and $w_a=0$.  
The Hubble parameter is then a derived parameter given by
$h = \sqrt{\omega_m/(1-\Omega_\Lambda)} = 0.73$ in the fiducial case.  

In the following, we first discuss the Fisher matrix for number counts
and clustering of clusters, before describing the calibration parameters
and CMB priors.  Throughout, we divide the redshift range into bins $l$ of
width $\Delta z = 0.05$.  Further, we bin clusters in logarithmic mass bins $m$
of width $\Delta \ln M =0.3$  from
the minimum mass $M_{\rm lim}(z)$ for each survey 
(\refsec{data}) up to a large cut-off
mass of $M_{\rm max} = 10^{16} M_{\odot}$. 
Since the mass limit varies with redshift,
the number of mass bins thus also varies somewhat across the redshift
range.  

\subsection{Number counts}
\label{sec:nc}
The FM for the number of clusters $N_{l,m}$ within the $l$-th redshift bin and $m$-th mass bin is 

\be
F_{\alpha\beta}=\sum_{l,m}\frac{\partial N_{l,m}}{p_\alpha}\frac{\partial N_{l,m}}{p_\beta}\frac{1}{N_{l,m}}
\label{eqn:ncfisher}
\ee

\noindent where the sum over $l$ and $m$ runs over intervals in the whole redshift range $z=0-1$ and cluster mass range $[M_{\rm lim}(z),\infty]$. 
We can write the abundance of clusters expected in a survey, within a given redshift and mass interval, using the mass function as:

\bea
&& N_{l,m} = \Delta \Omega \Delta z  \frac{d^2V}{dzd\Omega} 
\int_{M_{l,m}}^{M_{l,m+1}} dM^{\rm ob} \\ \nonumber
&& \ \ \ \ \ \ \ \int_0^{\infty} d\ln M\: n(M,z) p(M^{\rm ob}| M)
\eea

\noindent where $\Delta\Omega$ is the solid angle covered by the cluster 
survey, $\ln M_{l,m} = \ln M_{\rm lim}(z_l) + m \Delta\ln M$, and 
$n(M, z)$ is the mass function given in \refeq{nn}.  
Following~\citet{Lima2005}, we take into account the intrinsic scatter in the relation between true and observed mass, as inferred from a given mass proxy, by the factor $p(M^{\rm ob}| M)$ which is the probability for a given cluster mass 
with $M$ of having an observed mass $M^{\rm ob}$. Under the assumption of a log--normal distribution for the intrinsic scatter, with variance $\sigma^2_{\ln M}$, the probability is 

\be
 p(M^{\rm ob}| M)=\frac{\exp [-x^2(M^{\rm ob})]}{\sqrt{2\pi\sigma^2_{\ln M}}}
 \ee
 
 \noindent where
 
 \be
 x(M^{\rm ob})=\frac{\ln M^{\rm ob}-B_{\rm M}-\ln M}{\sqrt{2\sigma^2_{\ln M}}}.
 \label{eqn:scal}
 \ee

\noindent With these notations, we parameterize the $M^{\rm ob}-M$ relation, in addition to the intrinsic scatter,  by a systematic fractional mass bias $B_{\rm M}$.  With this prescription, the final expression for the number count FM is:

\bea
N_{l,m} &=& \frac{\Delta \Omega \Delta z}{2}  \frac{d^2V}{dzd\Omega} \label{eqn:dndzfm}\\
& & \times \int_0^{\infty}  d\ln M\: n(M,z) 
\left ({\rm erfc}[x_m]-{\rm erfc}[x_{m+1}] \right),  \nonumber
\eea
where erfc$(x)$ is the complementary error function.  

\subsection{Power spectrum}
\label{sec:ps}
We define the FM for the power spectrum of galaxy clusters as
\begin{align}
F_{\alpha\beta}=\frac{1}{(2\pi)^2}\sum_{m,n} \sum_{l,i} &
\frac{\partial \ln P^{mn}_{\rm h}(k_i,z_l)}{\partial p_\alpha}
\frac{\partial \ln P^{mn}_{\rm h}(k_i,z_l)}{\partial p_\beta}\vs
& \times V_{l,i}^{mn,\rm eff} k_i^2 \Delta k
\label{eqn:psfisher}
\end{align}

\noindent where the sum over $m,n$ runs over mass bins, while the sum in $l$ and $i$ runs over $z$ and $k$ intervals respectively. In what follows, we use  $k_{\rm max}=0.1\ h{\rm Mpc^{-1}}$ with $\Delta \log_{10} k=0.017$ for all surveys. Primordial NG of the local type modifies the shape of the power spectrum of galaxy clusters by introducing a scale-dependent bias on very large scales, therefore the choice of largest scale that can be probed by a cluster survey would significantly affect the constraints on $\fnl$. In practice, this is taken into account by introducing the window function of the survey in the calculation of $P(k)$. However, for simplicity, we approximate the influence of the window function by considering a cut off in the choice of $k_{\rm min}$. We use $k_{\rm min}=10^{-3}$ for the all-sky \planck\ survey, and $k_{\rm min}=10^{-2}$ for the partial sky surveys.

$P^{mn}_{\rm h}(k_i,z_l)$ is the cluster cross-power spectrum for mass bins $m$ and $n$, calculated for the given redshift and wavenumber through
\be
P^{mn}_{\rm h}(k_i, z_l) = b^m_{\rm eff}(z_l) b^n_{\rm eff}(z_l) P_L(k_i, z_l).
\ee
Here, $b^m_{\rm eff}$ is the mass function weighted effective bias, 
\be
b^m_{\rm eff}(z) = \frac{\int_{0}^{\infty} dM\:n(M,z)  b_L(M,z) ({\rm erfc}[x_m]- {\rm erfc}[x_{m+1}])}{\int_{0}^{\infty} dM\:n(M,z)  ({\rm erfc}[x_m]- {\rm erfc}[x_{m+1}])}.
 \label{eqn:beff}
\ee

\begin{figure}
  \begin{center}
       \includegraphics[width=80mm]{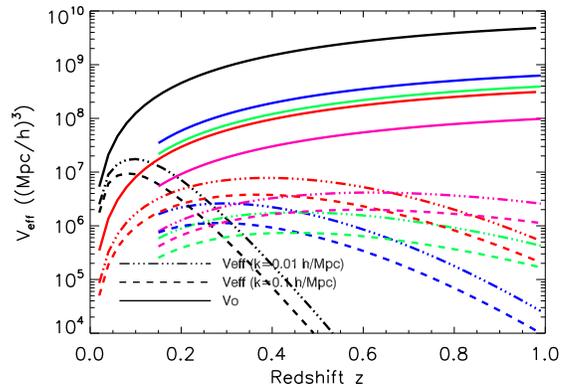} 
       \includegraphics[width=80mm]{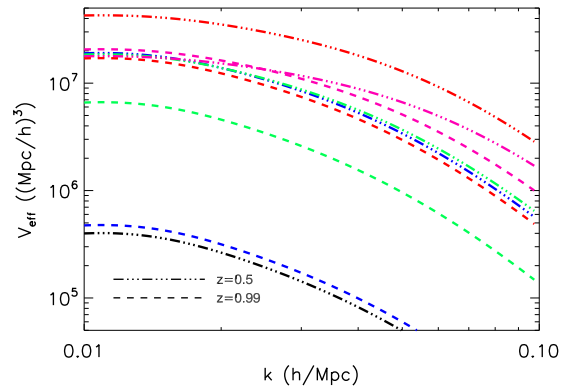}  
       \caption{The dependence on redshift {\it (top)} and wavenumber {\it (bottom)} of the effective volume (\refeq{veff}) for a single mass bin and each survey: \planck\ (black), SPT (green), SPTpol (magenta), ACTpol (blue), and CCAT--like (red). The effective volume is a weak function of wavenumber $k$ but strongly depends on the redshift.}
     \label{fig:veff}
  \end{center}
\end{figure}

 \noindent The effective
volume for mass bins $m,n$, wave number $k_i$, and redshift $z_l$ is given by
(see App. A of~\cite{Mak2012})
 
\begin{align}
 \frac{V^{mn,\rm eff}(k_i, z_l)}{V_0(z_l)} =\:& [P^{mn}(k_i, z_l)]^2 n_m(z_l) n_n(z_l) \label{eqn:veff} \\
& \times \Big[(n_m P^{mm} + 1)(n_n P^{nn} + 1) \vs
&\quad\   + n_m n_n (P^{nm} + \delta^{nm} n_m^{-1})^2\Big]^{-1},\nonumber
\end{align}

 \noindent where $V_0(z)$ is the comoving volume of the redshift slice $[z_l-0.01,z_l+0.01]$ covered by the given survey, and $n_m(z_l)$ is the cluster number density for mass bin $m$ at redshift $z_l$. 
The effective volume gives the weight carried by each bin in the $(z,k)$ space to the power spectrum Fisher matrix, and hence quantifies the amount of information contained in a given redshift- and $k$-bin.  
\reffig{veff} shows the redshift and scale dependence of the effective volume for the five cluster surveys. 
We find that $V_{\rm eff} \lesssim 0.3\,V_0$ for all redshifts and surveys
considered, even when not binning in mass, hence the cluster power spectrum is shot-noise dominated for all surveys.  As the lower panel of \reffig{veff} illustrates, Planck is most limited by shot noise, while SPTpol is least limited, as expected from their respective mass limits and coverage.  

\subsection{Calibration parameters}

In self-calibrating the true and observed cluster mass (\refeq{scal}), we introduce four nuisance parameters which specify the magnitude and redshift-dependence of the fractional mass bias $B_M(z)$ and the intrinsic scatter $\sigma_{\ln M}(z)$.  Following~\cite{Lima2005}, we assume the following parametrization:

\bea
& B_M(z)=B_{M0}(1+z)^\alpha \nonumber \\
& \sigma_{\ln M}=\sigma_{\ln M,0}(1+z)^\beta
\eea
Therefore the four nuisance parameters are $B_{M0}$, $\alpha$, $\sigma_{\ln M,0}$, and $\beta$.  
A negative value for $B_M$ corresponds to an underestimation of mass.  The mass bias accounts for the possibility of a systematic offset in the calibration of the observable mass scaling relation.  
We adopt fiducial values of $B_{M0}=0$, $\alpha=0$,  $\sigma_{\ln M}=0.1$,  $\beta=0$.  
In deriving the main results, we will not make any assumption on the four nuisance parameters and leave them free to vary.  
We will study the effect of assuming different priors on the four nuisance parameters on the $\fnl$ constraints in \refsec{nuisance}. 
 
 \subsection{CMB Prior}
In the following, we present results with the Fisher matrix for the \planck\ CMB temperature power spectrum $C_l$ added to the constraints from cluster counts and power spectrum. We calculate the full CMB fisher matrix with CAMB~\cite{Lewis2000} and method described in~\cite{Pritchard2008}. For the \planck\ experiment, we use the three frequency bands 100, 143 and 217 GHz, and the $C_l$ are calculated up to $l_{\rm max}=2500$. Our fiducial parameter set for the CMB experiment is, as described in the DETF report~\cite{Albrecht2006},  $\theta=(n_s, \Omega_b h^2, \Omega_\Lambda, \Omega_m h^2, w_0, A_s, \tau)$, where $A_s$ is the primordial amplitude of scalar perturbations and $\tau$ is the optical depth due to reionization. After marginalizing over the optical depth, we transform the Planck CMB fisher matrix to our cluster survey parameter set ${\theta}'=(n_s, \Omega_b h^2, \Omega_\Lambda, \Omega_m h^2, w_0, \sigma_8)$ by using the appropriate Jacobian matrix. The CMB imposes strong prior on the cosmological parameters. For example, $\Omega_m h^2$ is known to be measured with the CMB power spectrum to an exquisite precision, and this helps in  breaking parameter degeneracies in the constraints from cluster surveys. 
On the other hand, we note that the CMB power spectrum does not add constraints on $\fnl$ and therefore we compute the CMB fisher matrix for the Gaussian perturbation.

\section{Results}
\label{sec:result}

\begin{table*}
\caption{Marginalized $1\sigma$ errors on $\fnl$. The labels in the second column means the following. {\it standard}: The standard setup as indicated in~\refsec{nc} and~\refsec{ps}; {\it less conservative}: The conservative mass limit of the cluster survey is considered (see~\refsec{combine}; {\it photo-z}: Redshift binning correspond to the error of photometric redshift, i.e. $\Delta z=0.1$; {\it one mass bin}: No mass slicing in the fisher matrix, i.e. only one mass bin.  }
\begin{center}
\begin{tabular}{ll ccccc}
\hline\hline
Probes &  & Planck & ACTpol & SPT & SPTpol & CCAT--like\\
\hline
             dN/dz  &            standard       &     987 &           1473 &           2319 &           1463 & 815\\
              & less conservative          &  497 &           1069 &           1810 &           1230 & - \\
             &             photo-z          & 1540 &           1588 &           2355 &           1462 & 912\\
             &          one mass bin     &      $\infty$ &$\infty$ &          $\infty$ &          $\infty$ & $\infty$ \\
 \hline
              P(k)    &          standard    &         7 &             24 &             21 &             15 &13\\
              &    less conservative      &       4 &             16 &             17 &             13 & - \\
              &               photo-z            & 8 &             24 &             20 &             15 & 13\\
              &          one mass bin        &    160 &            272 &            267 &            124 & 115\\
\hline
       dN/dz + P(k)     &         standard         &    7 &             24 &             20 &             15&13\\
          &less conservative     &        4 &             16 &             16 &             13 & -\\
        &        photo-z        &     7 &             24 &             20 &             15 & 13\\
              & one mass bin     &       156 &            269 &            255 &            119 & 113\\
\hline
\end{tabular} 
\end{center}
\label{t:results}
\end{table*}

\subsection{Cluster counts and Power Spectrum}
\label{sec:ncandps}
The first four rows of \reftab{results} summarize the marginalized $\fnl$ constraints from $dN/dz$.  Under the standard setup: assuming we have spectroscopic redshifts $\Delta z=0.05$ and binning in cluster mass of $\Delta \log M=0.3$, the $\fnl$ constraints are $\sim10^{3}$. The relative constraining power of the different surveys can easily be interpreted by looking at $\Delta N$ shown in \refFig{Nz}. The Planck and CCAT--like survey, which have the largest $\Delta N$ at $z<0.3$ and $z>0.3$ respectively, similarly give the tightest constraints of  $\sigma_{\fnl}=987$ and $\sigma_{\fnl}=815$ respectively at the $68\%$ CL. The other three surveys are less stringent and the constraints are $>80\%$ worse than CCAT--like.

We study how the constraints change if we ignore the redshifts and consider combining information on all redshifts. With only one redshift bin, none of the surveys is able to constrain $\fnl$. This implies redshift information on each clusters is essential in constraining $\fnl$ using number counts.  So far we have assumed the optimistic scenario in which the cluster redshifts are spectroscopic. 
We relax this choice by using the photometric redshifts (i.e. $\Delta z=0.1$), this worsens the constraints, particularly for \planck\ in which $\sigma_{\fnl}$ increased by $59\%$ .

\begin{figure}
  \begin{center}
\includegraphics[width=0.55\textwidth]{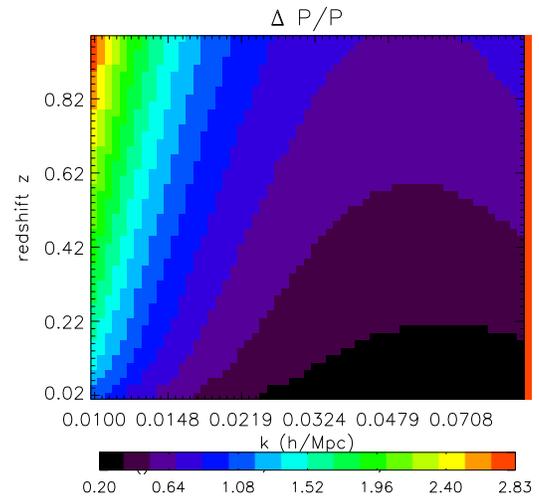}  
         \caption{Relative deviations of the non-gaussian power spectrum from the gaussian power spectrum, i.e. $\Delta P/P_{\rm G}$ for the \planck\ survey and $\fnl=100$. }
     \label{fig:deltap}
  \end{center}
\end{figure}

The constraints from power spectrum only are summarized in the row 5-9 in \reftab{results}. Overall the constraints are almost two orders of magnitude better than that from number counts only. This is because the scale dependence of the halo bias is very sensitive to non-gaussianity (e.g.~\cite{Scoccimarro2004,Matarrese2008,Dalal2008}). The best constraint come from the all-sky \planck\ survey, with $\sigma_{\fnl}=7$. This result is largely due to its large sky coverage ($f_{\rm sky}\approx0.7$). It was shown in previous studies~\cite[e.g.][]{Loverde2008,Grossi2009} that the deviation of the halo bias, and hence the power spectrum, between non-gaussian and gaussian case is most prominent at large scale. We illustrate this in \reffig{deltap} with the $\Delta P/P$ in different redshift and wavelength bins. The survey that probes longer wavelength modes, as in the case of \planck\ with $k_{\rm min}=10^{-3}$, therefore has the highest sensitivity to $\fnl$. 

\begin{figure}
  \begin{center}
\includegraphics[width=0.40\textwidth]{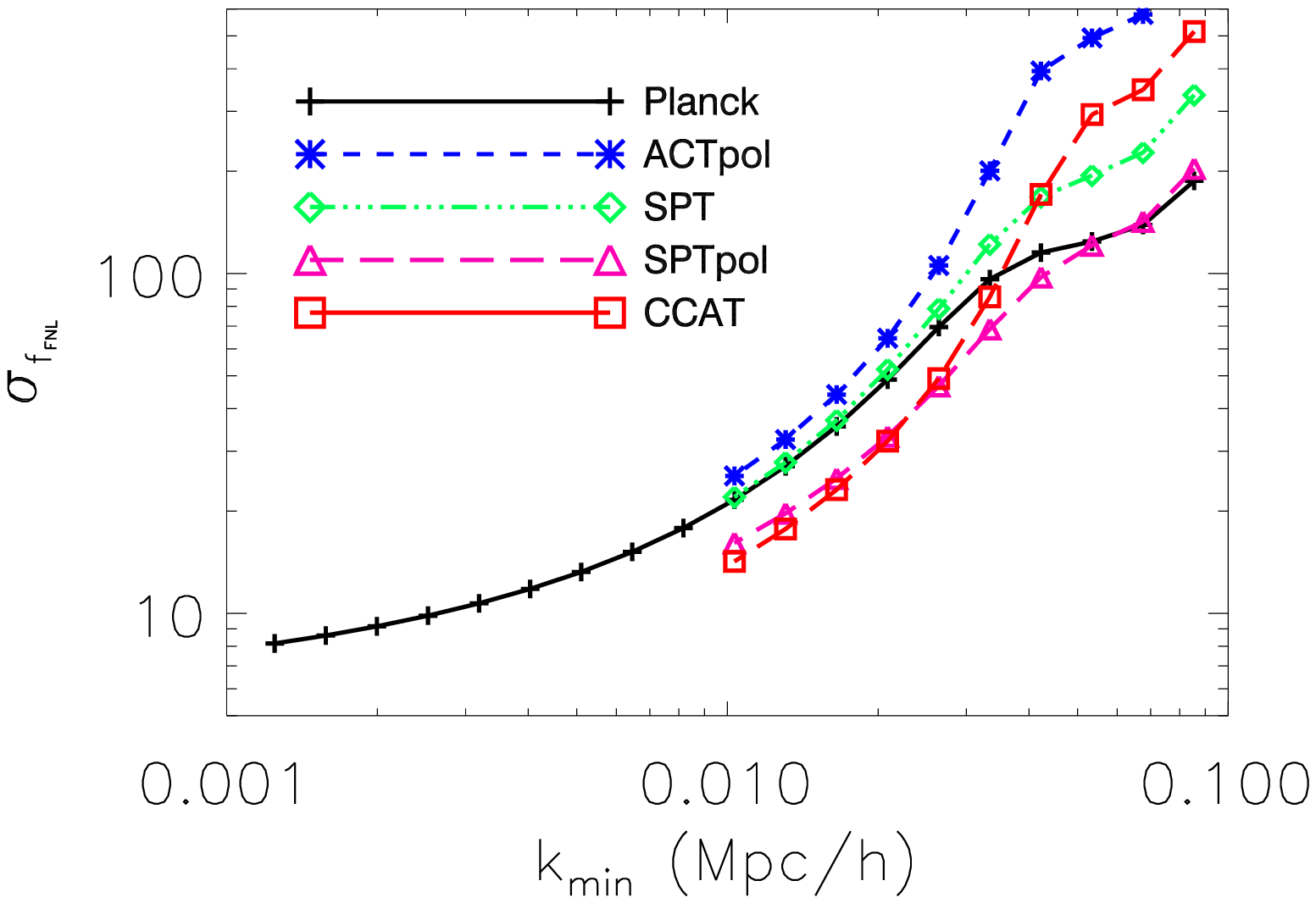}  
\includegraphics[width=0.40\textwidth]{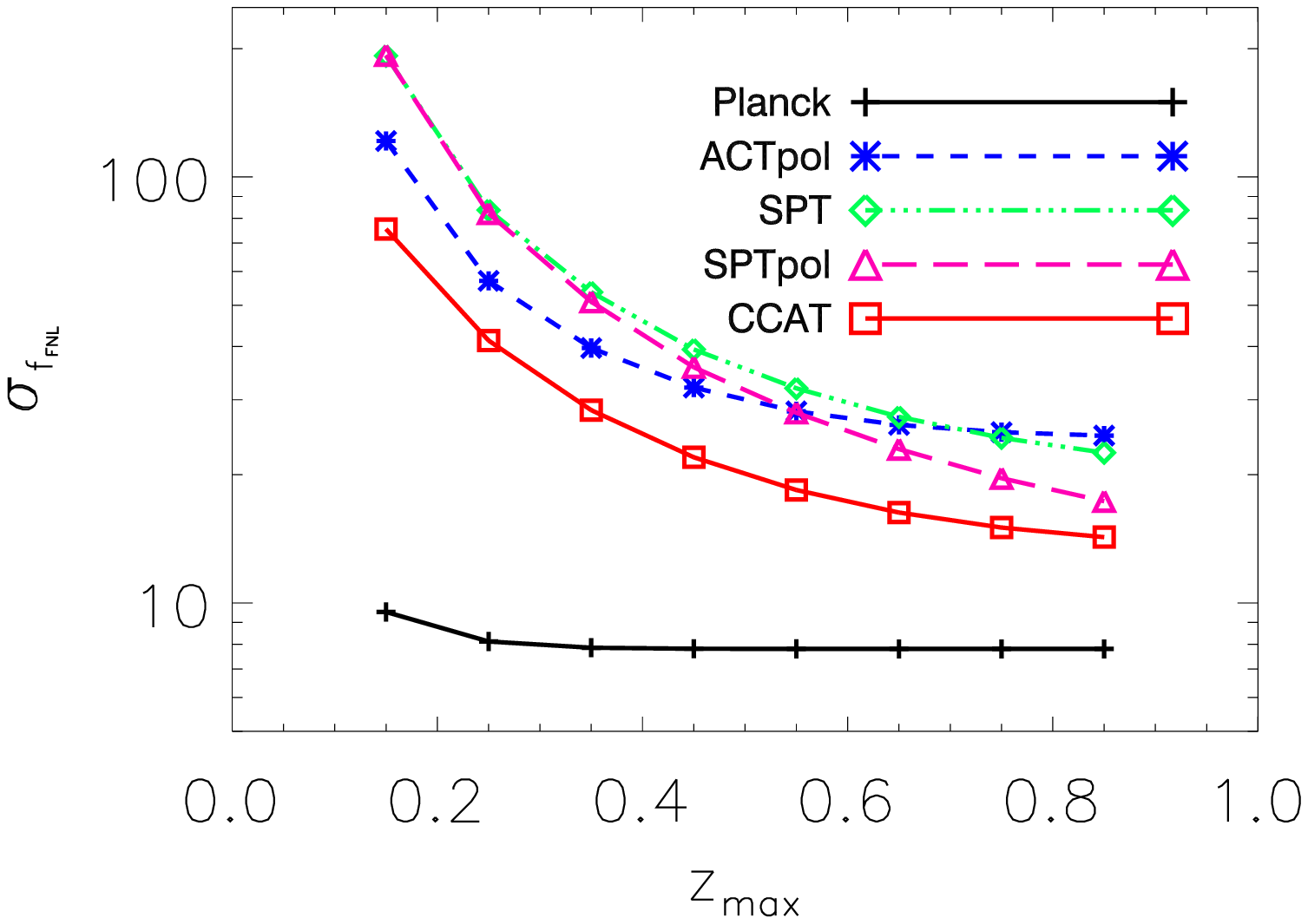}  
         \caption{Fully marginalized constraints on $\fnl$ from the power spectrum of clusters only, as a function of minimum wavenumber $k_{\rm min}$ (upper) and maximum cluster redshift $z_{\rm max}$. (lower).}
     \label{fig:sigmakz}
  \end{center}
\end{figure}

We here quantify the sensitivity of the non-gaussianity constraints to the adopted $k_{\rm min}$ value in the analysis of the power spectrum. We address this by computing $\sigma_{\fnl}$ as a function of $k_{\rm min}$, as shown in \reffig{sigmakz}. As expected, the constraints significantly improve when we consider smaller $k$ values. This is partly because more information is included in the fisher matrix and, more importantly, the effect of NG on halo bias is most prominent at the largest scales. It is interesting to note that if we limit $k_{\rm min}$ to $10^{-2}$ Mpc/h also for the \planck\ survey, then the derived $\fnl$ constraint would be similar to, and even slightly worse than the one derived from the CCAT--like and SPTpol survey.

It is worthwhile to note that the effective volume $V_{\rm eff}$ also impacts the $\fnl$ constraints by its $z$ and $k$ dependence on the fisher matrix, as shown in \reffig{veff}. This is particularly useful in understanding the relative merits of the four partial sky surveys that have similar survey area and $k$-range. Indeed the relative constraining power of these four surveys is reflected in the redshift distributions of the effective volume: while CCAT--like and SPTpol has the largest $V_{\rm eff}$ at $z<0.7$ and $z>0.7$ respectively (upper panel of~\reffig{veff}), their errors on $\fnl$ are similar and are $40\%$ smaller than ACTpol and SPT. To further understand which redshift range contributes most to the $\fnl$ constraints, we show in \reffig{sigmakz} the $\fnl$ constraints as a function of the maximum cluster redshifts. 
If we limit the cluster samples to $z\le0.6$, both CCAT--like and ACTpol survey give tighter constraint than SPTpol and SPT. As we extend to larger maximum redshift, SPTpol survey then gives tighter constraint than SPT and ACTpol survey as its $V_{\rm eff}(z)$ is decreasing at a smaller rate. This result shows that the constraints leverage on clusters in disjoint redshift ranges and these surveys provide complementary information on $\fnl$ constraints from power spectrum. 

\subsection{Combined constraints}
\label{sec:combine}
The constraints on $\fnl$ when combining both number counts and power spectrum are summarized in rows 10-13 of \reftab{results}. While combining the two probes helps to break degeneracies and improves constraints on cosmological parameters, the constraint on $\fnl$ does not improve. The constraining power is mainly driven by the power spectrum. We note that the best constraint estimated here ($\sigma_{\fnl}=7$ from \planck\ cluster survey) is a factor of 2 (when accounting the difference in $\fnl^{\rm LSS}$ and $\fnl^{\rm CMB}$) better than the current upper limits as measured by CMB experiments (e.g. $32\pm21$~\cite{Komatsu2011}) and a factor of 14 better than large scale structure probes (e.g. $\Delta \fnl=\pm96$~\cite{Slosar2008}). In the long run, however, constraints from new CMB measurements from \planck\ would push the upper limits further down to $\sigma_{\fnl}<5$~\cite{Yadav2007}. These Planck limits also consider polarization maps and disregarding potential foregrounds, thus it these CMB limits should be regarded as the optimistic constraint. Next generation all-sky X-ray surveys~\cite{Sartoris2012,Pillepich2012}$ would achieve \sigma_{\fnl}\approx10$. Note that constraints from CMB are based on different physics than large scale structures and suffered from different systematics. The SZ clusters surveys therefore provide a useful complement to the $\fnl$ information we can derive from the CMB maps. 

\begin{figure}
  \begin{center}
\includegraphics[width=0.40\textwidth]{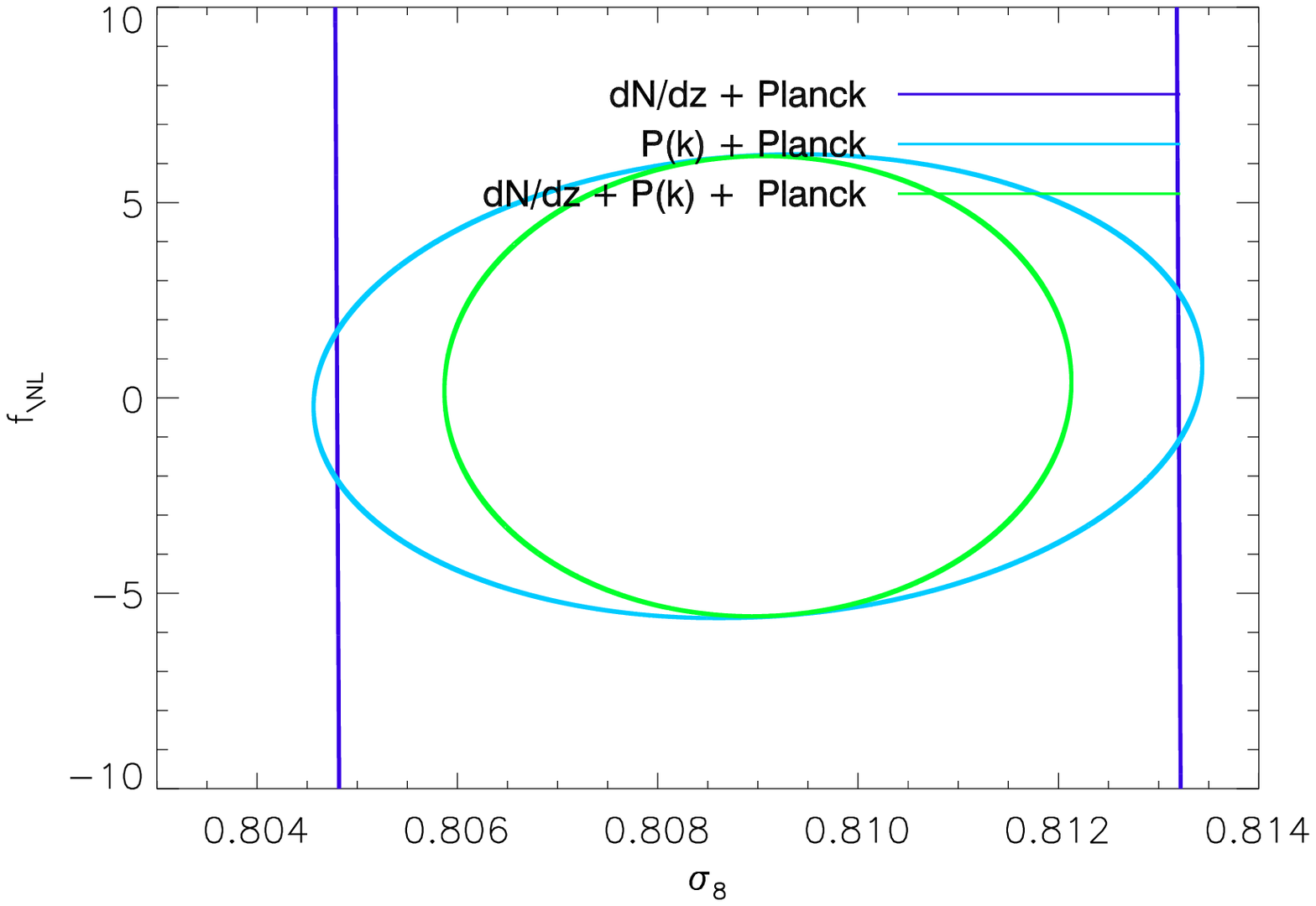}
\includegraphics[width=0.40\textwidth]{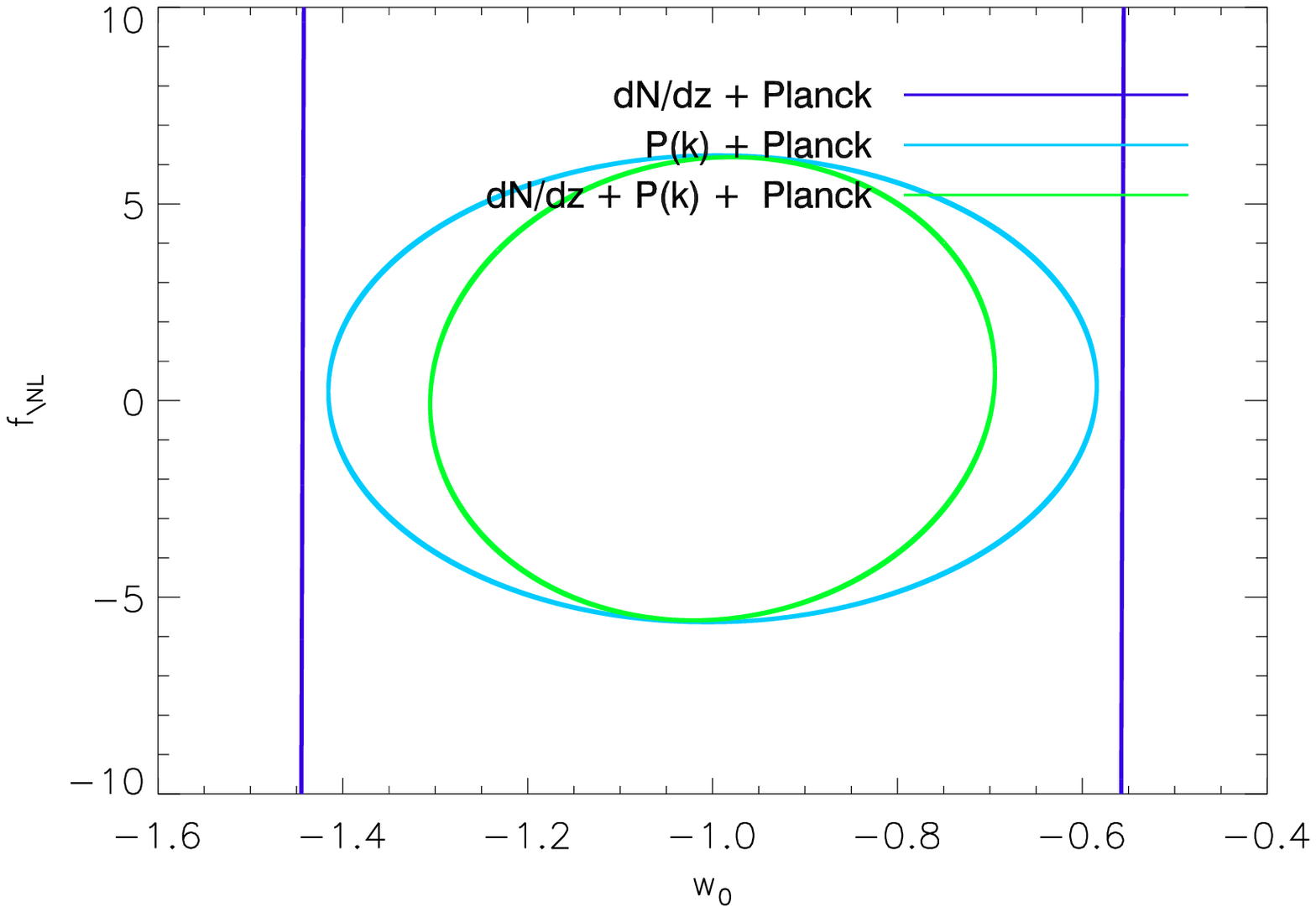}
\includegraphics[width=0.40\textwidth]{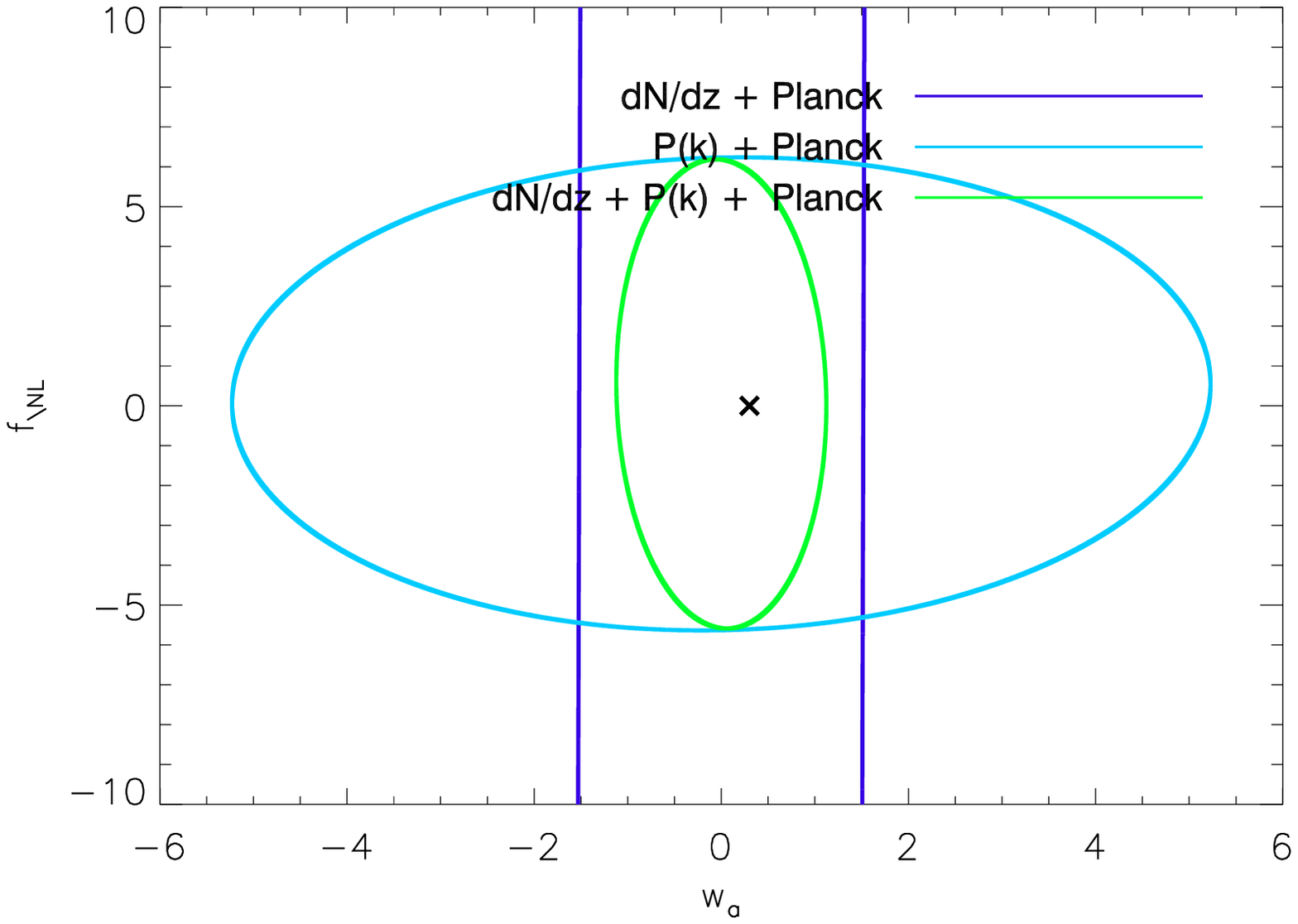}
       \caption{Joint constraints on the $\fnl$ and (counterclockwise from top left) $\sigma_8$, $w_o$, and $w_a$. All curves denote $68\%$ confidence level, and are for number counts only (blue), power spectrum only (cyan), and combination of the two (green). The Planck CMB power spectrum priors are assumed. }
     \label{fig:contour}
  \end{center}
\end{figure}

\reffig{contour} illustrates the most important degeneracies of $\fnl$ with cosmological parameters ($\sigma_8$, $w_0$, $w_a$) for the \planck\ survey. These plots clearly demonstrate the complementarity that the number counts and clustering have to constrain $\fnl$ and standard cosmological parameters, particularly the redshift dependence part of equation of state of dark energy $w_a$. It is obvious that the $\fnl$ is almost non-degenerate with other cosmological parameters. Only mild degeneracies are seen in the joint constraint with $w_0$ and $w_a$. Therefore the constraints on other parameters have little effect on $\fnl$.

It is important to keep in mind that these results are based on a conservative mass limits, i.e. clusters are expected to be detected with $S/N\ge5$ for all surveys. We examine improvements in the $\fnl$ constraints when using more optimistic mass limits for each survey according to what is outlined in \refsec{data}. Using number counts only, the constraint from \planck\ is improved by $50\%$, while constraints from other surveys are only slightly improved (by $16-27\%$). However, these constraints are worse than those from power spectrum. Using power spectrum only, the constraints are improved only marginally for SPTpol and SPT, but largely by $30-40\%$ for \planck\ and ACTpol. It is interesting to notice that ACTpol is now slightly better in constraining $\fnl$ than SPT while it was the opposite case when using the conservative mass limits. This is mainly due to the larger increase in detected clusters for the ACTpol survey in the optimistic mass limits.  

As a worse case scenario, we also consider the case when there is no information on the cluster mass, i.e. one mass bin. For all surveys, the combined constraints are significantly worsened by a factor of $>10$. Yet, the CCAT--like and SPTpol constraints with single mass bins are still comparable to current upper limits.

\section{Discussion}
\label{sec:discussion}
\subsection{Uncertainties in scatter of mass observable relations}
\label{sec:nuisance}
Our forecasts of $\fnl$ assumed no prior on the nuisance parameters in the mass-observable relations: $B_{M,0}$, $\sigma_{M,0}$, $\alpha$, $\beta$. However, allowing more freedom to the scaling relations parameters results in degradation of the final constraint. In practice, we can expect some external constraints on these parameters by using detailed studies of individual clusters or combining different information from optical, lensing, X-ray and SZ measurements. To better understand the importance of self-calibration of systematic uncertainties on the non-Gaussianity constraints, we repeat the forecasts on different priors on the four nuisance parameters.

The results are summarized in \reftab{result_prior}. We have used the current knowledge on the calibration on the mass proxies from X-ray and lensing measurements~\cite[e.g.][]{Zhang2010} and assume priors on the nuisance parameters of $\Delta \sigma_{M,0}=0.1$, $\Delta \beta=1$, $\Delta B_{M,0}=0.05$, and $\Delta\alpha=1$. We refer this to the weak prior. Furthermore, we refer to the case strong prior when we assume that all these four parameters are held fixed at their fiducial values. We find that the $\fnl$ constraints only slightly improves when using number counts only and has negligible improvement in the joint constraints when considering the weak prior. In the case of strong prior, the constraints are significantly tightened by a factor of $4-8$ from number count only, but negligibly from power spectrum only. This results in appreciable improvements in the joint constraints for the partial sky surveys but not for \planck, as measured from the cluster. 

As the $\fnl$ joint constraint is essentially driven by the power spectrum, we conclude that it is largely insensitive to priors on nuisance parameters. This is consistent with the scenario shown in~\cite{Sartoris2012} as they calculated the ratio of non-Gaussian to Gaussian at a wide range of values of the mass bias $B_{M0}$ and scatter $\sigma_{M0}$. They showed that the ratio on number counts change by less than $0.1\%$, and is negligible on the effective bias except at the very large scales ($40\%$ at $k\approx10^{-3} {\rm Mpc^{-1}}$). This justifies the weak dependence of the $\fnl$ constraints on prior since the choice of prior affects more the prediction for the cluster counts on $\fnl$, thus show smaller dependence than that from clustering measurements.

\begin{table*}
\caption{Fractional improvement $\frac{\sigma_{\fnl, no}}{\sigma_{\fnl, weak/strong}}$, with various priors (see~\refsec{nuisance}). }
\begin{center}
\begin{tabular}{ll ccccc}
\hline\hline
Probes & prior & Planck & ACTpol & SPT & SPTpol & CCAT--like\\
\hline
          dN/dz  &         weak  & 1.63  & 1.54  & 2.13  & 2.22 & 1.63\\
              &     strong  & 7.68  & 3.48  & 5.41  & 4.39 & 4.39 \\
\hline
           P(k)        &   weak  & 1.00  & 1.01  & 1.01  & 1.01 & 1.01 \\
                  & strong  & 1.02  & 1.04  & 1.01  & 1.02 & 1.04 \\
\hline
     dN/dz+P(k)    &       weak  & 1.01  & 1.02  & 1.01  & 1.01 & 1.01 \\
         &   strong  & 1.66  & 3.46  & 3.15  & 2.80 & 3.03 \\
 \hline
\end{tabular} 
\end{center}
\label{t:result_prior}
\end{table*}


\subsection{Comparison to previous work}
Forecast of the local-type non-gaussianity was previously done with galaxy cluster probes in other wavelengths, e.g. future X-ray surveys (\cite{Sartoris2012} and \cite{Pillepich2012}) and optical surveys (\cite{Fedeli2011}, \cite{Cunha2010}, \cite{Oguri2009}). This study is the first attempt in exploring the $\fnl$ constraints from the SZ cluster survey and our results are in broad agreement with these previous studies. A detailed comparison between our work and these studies is not straightforward because of the different survey specifications, prescriptions for the mass function and halo bias, and the type of cluster probes used. The more interesting comparison can be made with the results from~\cite{Sartoris2012} and~\cite{Pillepich2012} who put constraints with cluster number counts and clustering and similar subsamples. 

In terms of cluster sample ever considered by other authors, the subsample from eRosita~\cite{Pillepich2012} of the most massive 1000 clusters ($M_{\rm 500,crit}\ge2.2\times10^{14} h^{-1}M_\odot$ at $z\ge1$, "magnificent 1000") is the most similar to the \planck\ cluster sample. This is because the limiting mass of the \planck\ catalog behaves in a more similar way to the X-ray catalog due to its large beams that significantly smoothes the signal, especially when the angular size of the cluster is small. We forecast a tighter constraint ($\sigma_{\fnl}=7$) than theirs ($\sigma_{\fnl}=26$), when both assuming photometric redshift information and adding the Planck CMB prior. One should keep in mind that the non-gaussianity effect is more prominent at high redshift, so the "magnificent 1000" which contains cluster at $z\ge1$ should in principle be more sensitive to $\fnl$ constraint. The discrepancies could probably be due to a number of differences in the analysis: (a) they considered different types and number of nuisance parameters that used to model the uncertainties in the scaling relations, such as $L_{\rm X}-M$, $T_{\rm X}-M$. This is a conservative approach but at the same time degrade their constraints considerably. (b) the mass--redshift distribution of the X--ray sample are different from our SZ samples. The SZ samples tend to have higher fraction of massive clusters, i.e. the regime where the NG effects are more prominent. This makes the SZ clusters more sensitive to the $\fnl$ constraints.

In terms of cluster probes, we use the same ones in~\cite{Sartoris2012}, except that we additionally consider slicing the $P(k)$ in mass bins. This makes our constraint comparable to theirs even if our samples have much lower statistics ($\sigma_{\fnl}=7$ for $\approx1000$ clusters in the \planck\ survey vs $\sigma_{\fnl}=11$ for $\sim10^{6}$ clusters in the WFTX Wide survey). On the other hand, if we do not consider mass bins in $P(k)$ then our constraint degrade to $\sigma_{\fnl}=156$, which is an order of magnitude worse. The improvement mainly comes from the mass dependence of the halo bias which in turn provides extra information of the shape of the power spectrum. This suggests the slicing in mass bins compensates for the poor statistics.

\section{Conclusions}
\label{sec:conclusion}
In this work, we exploited the large cluster samples expected from current and upcoming SZ surveys to place constraints on primordial non-Gaussianity of the local type. Making use of the cluster number counts and power spectrum, and taking into account the self-calibration of mass-observable scaling relations, we employed the Fisher matrix analysis to forecast the sensitivities of various SZ surveys in constraining the $\fnl$ parameter.

The main results are presented in \refsec{result}. We find that the induced scale-dependence of halo bias (through a term that is proportional to $k^{-2}$) by local type non-Gaussianity provides a very effective way to put strong constraints on $\fnl$. This makes the power spectrum a more powerful probe than number counts, by at least two orders of magnitudes. As a result, all--sky surveys such as \planck, which probe better the most NG sensitive regime at large scales, are more favorable to measure $\fnl$. The best constraint we obtain is from Planck (combined constraints, $\sigma_{\fnl}=7$) and is mainly driven by the power spectrum. The partial--sky surveys SPT, SPTpol, ACTpol, and CCAT--like, however, are a factor of $>2$ less constraining than \planck\, with $\sigma_{\fnl}=13-24$. The best constraint we obtained, which is based on conservative assumptions on uncertainties in mass-observable relations and mass thresholds, is a factor 2 better than that measured from WMAP CMB and is comparable to the expected results from the Planck CMB non-Gaussianity studies ($\fnl\le5$). The SZ cluster surveys therefore provide a useful complement to the $\fnl$ information we can derive from CMB maps. However one should note that these results are based on different physics and suffering from different systematics than the probes considered in this work. We also show that $\fnl$ have little degeneracy with other cosmological parameters and it is only mildly degenerate with $w_0$ and $w_a$. Thus the constraints on other parameters have little effect on $\fnl$.

We investigate the sensitivity of our results to various aspects of survey specification in \refsec{discussion}. We find that the errors on $\fnl$ are mainly driven by the power spectrum and therefore they are insensitive to priors on nuisance parameters. Only when we have perfect knowledge on the uncertainties of these parameter we can improve the constraint by a factor of 1.7-3.5, with \planck\ being the least benefited.  This, however, requires very good understanding of the mass calibration which may be possible with multi--frequency followup measurements. In addition, the cluster selection function plays a less significant role in which $\sigma_{\fnl}$ is marginally improved when we consider more optimistic mass limits.

Our results are in broad agreement with previous studies that explores the constraining power of $\fnl$ from X--ray and optical cluster surveys. The values of $\sigma_{\fnl}$ we obtained are comparable to the ones quoted in these other works even if the SZ samples contain at least two orders of magnitude less clusters. This is because, unlike previous authors, we considered mass slicing in power spectrum when calculating the fisher matrix and this greatly improves the constraints. This suggests the slicing in mass bins compensates the poor statistics of the SZ cluster samples. We realize that the SZ surveys we considered here are very promising and their cluster catalogues would be available in the very near future, since most of them are operating. Therefore, our $\fnl$ forecasts can readily be referenced.

\acknowledgments
We would like to thank Keith Vanderlinde for useful discussion of the sensitivity of the CCAT cluster survey.
EP acknowledges support from  JPL-Planck subcontract 1290790.
DM acknowledges support from USC Stauffer Fellowship. 
EP and DM were partially supported by  NASA grant NNX07AH59G.

\bibliography{ms}
\end{document}